# Narrowing band gap chemically and physically: Conductive dense hydrocarbon

Takeshi Nakagawa[a,*], Caoshun Zhang[a], Kejun Bu[a], Philip Dalladay-Simpson[a], Martina Vrankić[b], Sarah Bolton[c], Dominique Laniel[c], Dong Wang[a], Akun Liang[c], Hirofumi Ishii[d], Nozomu Hiraoka[d], Gaston Garbarino[e], Angelika D. Rosa[e], Qingyang Hu[a], Xujie Lü[a], Ho-kwang Mao[a,f], and Yang Ding[a,**]

[a] Center for High Pressure Science and Technology Advanced Research, 10 Xibeiwang East Road, Haidian District, Beijing, 100094, P.R. China
[b] Division of Materials Physics, Ruđer Bošković Institute, Bijenička 54, 10000 Zagreb, Croatia
[c] Centre for Science at Extreme Conditions and School of Physics and Astronomy, University of Edinburgh, EH9 3FD Edinburgh, United Kingdom
[d] National Synchrotron Radiation Research Center (NSRRC), Hsinchu 30076, Taiwan, R. O. C.
[e] European Synchrotron Radiation Facility (ESRF), 71 avenue des Martyrs, 38000 Grenoble, France
[f] Shanghai Key Laboratory of Material Frontiers Research in Extreme Environments (MFree), Shanghai Advanced Research in Physical Sciences (SHARPS), Shanghai 201203, P. R. China

* takeshi.nakagawa@hpstar.ac.cn; ** yang.ding@hpstar.ac.cn (corresponding authors)

**Abstract:** Band gap energy of an organic molecule can be reduced by intermolecular interaction enhancement, and thus, certain polycyclic aromatic hydrocarbons (PAHs), which are insulators with wide band gaps, are expected to undergo insulator-metal transitions by simple compression. Such a pressure-induced electronic transition can be exploited to transform non-metallic organic materials into states featuring intriguing electronic characteristics such as high-temperature superconductivity. Numerous attempts have been made to metalize various small PAHs, but so far only pressure-induced amorphization well below the megabar region was observed. The wide band gap energy of the small PAHs and low chemical stability under simple compression are the bottlenecks. We have investigated the band gap energy evolution and the crystal structural compression of the large PAH molecules, where the band gap energy is significantly reduced by increasing the number of π-electrons and improved chemical stability with fully benzenoid molecular structure. Herein, we present a pressure-induced transition in dicoronylene, $C_{48}H_{20}$, an insulator at ambient conditions that transforms into a semi-metallic state above 23.0 GPa with a three-order-of-magnitude reduction in resistivity. In-situ UV-visible absorption, transport property measurement, Raman spectroscopy, X-ray diffraction and density functional theory calculations were performed to provide tentative explanations to the alterations in its electronic structure at high pressure. The discovery of an electronic transition at pressures well below the megabar is a promising step towards realization of a single component purely hydrocarbon molecular metal in the near future.

**Keywords:** Polycyclic aromatic hydrocarbon; chemical compression; high pressure; insulator-to-semiconductor transition; phase transition

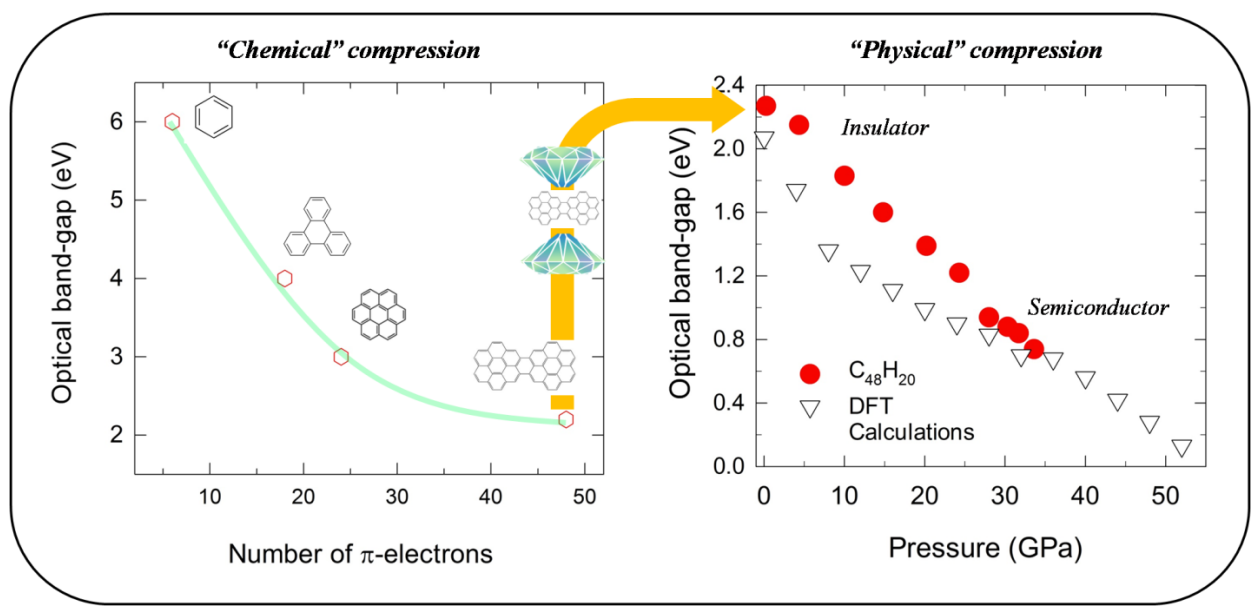

Table of contents

# Introduction

Insulator-to-metal (I-M) transition has been a topic of fundamental interest ever since the introduction of a model for the electronic band structure of solids.[1] Back in 1968, Ashcroft pointed out that metallic hydrogen would be a room temperature (RT) superconductor,[2] and since then the realization of metallic hydrogen has been a long-standing challenge in condensed matter physics. However, experimental confirmation of metallic hydrogen is extremely challenging, which requires ultra-high pressures of at least 350 GPa.[3] With the rapid development of experimental high-pressure technology, some researchers have claimed the existence of possible metallic and semi-metallic hydrogen,[4,5,6,7] but an unambiguous experimental confirmation of such observations remains extremely difficult. More recently, efforts have focused on hydrogen-dominant alloys such as binary and ternary metal hydrides and superhydrides, in which the electron density on the hydrogen atoms is pre-compressed, comparable to pure hydrogen compressed to megabar pressures.[8,9,10,11] In these compounds, the hydrogen atoms are expected, at much lower pressure, to mimic the properties of idealized pure metallic hydrogen.

Along the same line, organic molecular compounds composed of light elements such as carbon are very promising for superconductivity at high critical temperature ($T_c$). Theoretically, materials abundant in carbon and hydrogen can provide the necessary high frequencies in the phonon spectrum and a robust electron-phonon interaction. Specifically, much research has been conducted on polycyclic aromatic hydrocarbon (PAH) molecules, which are materials abundant in both hydrogen and carbon, in the pursuit of high $T_c$ materials.[12,13,14] In particular, Ashcroft highlighted in his theoretical research that there exists a narrow pressure range between 180-200 GPa where benzene ($C_6H_6$) may exhibit a molecular metallic state. The concept involves the synthesis of hydrogen and carbon "alloy" that can effectively decrease its band gap at RT and potentially transition into a metallic state under modest pressure.[15] Similarly, theoretical calculations on triphenylene ($C_{18}H_{12}$) have shown that it becomes metallic at 180 GPa.[16] Despite several experimental attempts to discover a metallic molecular state in a carbon-hydrogen system, the only PAH molecule that has been reported to display an I-M transition is pentacene ($C_{22}H_{14}$), observed at 27 GPa by Drickamer in 1962.[17] However, this phenomenon has not been reproduced, as other studies found pentacene to undergo amorphization beyond 11 GPa.[18,19,20] Other attempts to discover a metallic state in PAHs have only confirmed the occurrence of an irreversible chemical reaction at 8 – 20 GPa,[21,22,23] which is far below the expected critical pressure for the formation of a metallic state.

To realize a conducting state in hydrocarbons, band gap energy at ambient condition must be further pre-compressed and improve chemical stability of the molecule, which allows band gap to be closed with physical compression before the molecule decomposes. It is known that the band gap energy of PAH molecules can be tuned effectively by increasing the number of aromatic rings in PAH molecules, or more specifically, the number of π-electrons.[24] For instance, there are six π-electrons in benzene which features an optical band gap of approximately 6 eV,[25] while there are 18 π-electrons in triphenylene, reducing the optical band gap to 4 eV,[24] and 2.8 eV for coronene with 24 π-electrons (Supplementary Fig. 1).[26] This is due to the condensation of highly conjugated π-systems. Furthermore, our recent high-pressure studies of large PAH molecules in which the benzene rings are fused in a disc-like fashion have shown that they can be chemically stable even at pressures above 20 GPa.[26] These findings imply that large PAH molecules have a true potential for realizing the I-M transition and possible superconductivity, at high pressure.

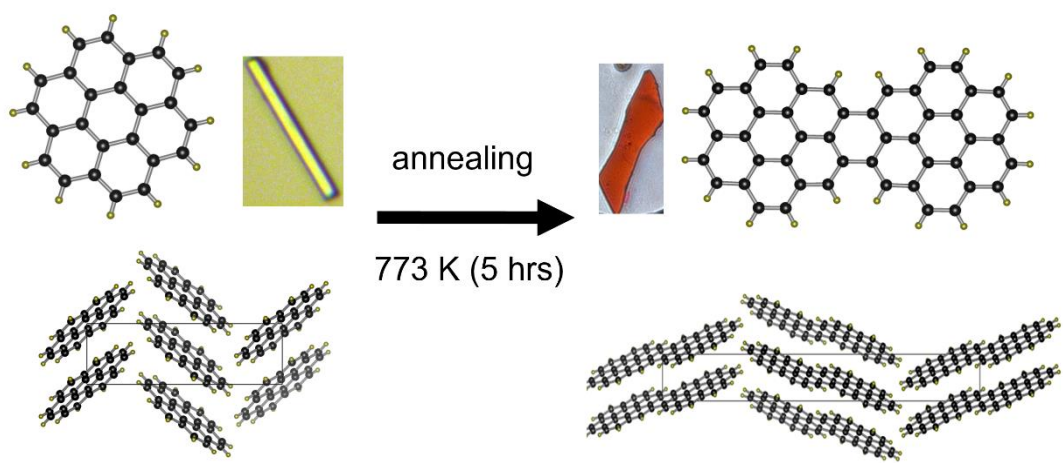

**Scheme 1.** Dicoronylene ($C_{48}H_{20}$) obtained via vapor phase fusion reaction of coronene ($C_{24}H_{12}$) molecules. Molecular structures, crystal structures and optical micrographs of $C_{24}H_{12}$ and $C_{48}H_{20}$.

Herein, we focus on the effect of pressure on the large PAH molecule, dicoronylene ($C_{48}H_{20}$), obtained by chemical fusion of two coronene molecules results in an optically red single crystal. (Scheme 1) The research utilizes various techniques including optical absorption, electrical resistance measurements, Raman and IR spectroscopy, powder and single crystal X-ray diffraction, and *ab initio* calculations supporting the experimental data. At ambient conditions, the band gap energy of $C_{48}H_{20}$ is reduced to 2.21 eV from 2.8 eV for coronene monomer. We observed continuous band gap closure from 2.21 eV at ambient conditions down to 0.7 eV at 33.6 GPa, accompanied by crystal optical color changes from red to complete black. We discovered the semi-metallic nature of $C_{48}H_{20}$ emerges in a pressure range between 23.0-38.0 GPa, which was confirmed by temperature-dependent electrical resistance measurements. Furthermore, the crystal structure transition was observed using XRD measurements and Raman spectroscopy and its reversibility was investigated using IR spectroscopy.

In particular, synchrotron single-crystal X-ray diffraction (SCXRD) revealed that the semi-metallic state emerges when the intermolecular nearest-neighbor C-C distance is shorter than 2.8 Å and irreversible chemical reaction take place at C-C distance of 2.6 Å. To the best of our knowledge, no other single-crystal data on PAHs have yet been reported to investigate structure-to-properties relations. The first-principles calculations shed light on the chemical bonding and electronic properties and corroborate the experimental findings, where the band gap closure under physical pressure occurs at 28.0 GPa. We have demonstrated that the increasing number of π-electrons can significantly chemically pre-compress the band gap, from which physical compression can further reduce the band gap, leading to a transition from insulator to semiconductor. The discovery of an electronic transition at pressures far below one megabar is a promising step towards the realization of a purely organic single-component molecular metal in the near future.

# Results and discussion

### Characterization at ambient pressure

Details of the $C_{48}H_{20}$ preparation can be found in our previous publication.[27] Single-crystal XRD (SCXRD) of $C_{48}H_{20}$ under ambient conditions shows that it crystallizes in a monoclinic *β*-herringbone assembly with the centrosymmetric space group $P2_1/c$, whose structure matched those of Goddard et al. (1995).[28] (Refinement result: $a = 10.3940(8)$ Å, $b = 3.8402(3)$ Å, $c = 31.990(3)$ Å, $β = 90.248(2)°$, $V = 1276.89(17)$ Å$^3$. The final anisotropic full-matrix refinement converged at $R_1 = 2.4$ % for the observed data and $_wR_2 = 7.4\%$ for all data). Details of the crystal structure extracted from the refined SCXRD data can be found in the Supplementary information Table S1. The crystalline $C_{48}H_{20}$ exhibits strong red fluorescence emission bands when irradiated with lasers emitting in visible frequencies. The weak Raman modes recorded with laser sources with wavelengths of 633, and 532 nm are therefore, obscured by the very high fluorescence. Out of the 204 active Raman modes of $C_{48}H_{20}$, only 20 lines were observed when the laser excitation was at the wavelength of 1064 nm. The optical band gap of $C_{48}H_{20}$ at ambient conditions was estimated to be 2.21 eV by extrapolating the linear portion of the Kubelka-Munk function plots[29,30] of the UV-vis absorption spectra. This value is much smaller compared to the other PAHs, for example, benzene (6 eV), naphthalene (4.2 eV), phenanthrene (4.16 eV), triphenylene (3.95 eV), tetracene (2.54 eV), pyrene (3.03 eV), and coronene (2.8 eV).[24,25,26,31,32] The pressure-dependent electronic structures at various pressure points were calculated using density functional theory (DFT) based on the ambient pressure structure to predict the pressure required for the band gap closure. The band gap closure for $C_{48}H_{20}$ was estimated to occur at around 50.0 GPa, which is still very high for organic molecules, but undoubtedly reachable, especially compared to predictions for other PAH molecules, e.g., 190 GPa for benzene, 160 GPa for triphenylene, and 117 GPa for benza[a]anthracene.[14,16,33]

## Optical absorption spectroscopy and band gap closure

Figure 1a shows the spectra obtained from in-situ UV-vis absorption measurements. Upon compression, the absorption increases and broadens, corresponding to the change in crystal optical color from red to dark red, opaque black above 10.0 GPa, and completely black above 24.0 GPa (Fig. 1b). The transformation of this crystal color is reversible, with the original crystal color restored when the applied pressure was completely released from 14.0 GPa (Supplementary Fig. 2a), but irreversible when released from 60.0 GPa (Supplementary Fig. 2b). At higher pressures, a broadening and continuous red-shift of the UV-Vis absorption spectra extends to longer wavelengths and eventually covers the entire visible and near-infrared (NIR) range. Above 24.0 GPa, the absorption edge shows a discontinuous, sharp red-shift, indicating a structural transition accompanied by a complete blackening of the crystal, where it absorbs all visible and NIR light, i.e., 0 % transmittance. Figure 1c shows the optical band gap as a function of pressure. The optical band gap of $C_{48}H_{20}$ extracted from the absorption spectra exhibits systematical decrease from 2.2 eV at ambient conditions to 1.2 eV at 24.0 GPa, and a discontinuous jump to 0.9 eV was observed at 26.6 GPa. When the applied pressure was further increased above 30.0 GPa, the spectra continued to show a red-shift, but it was not possible to estimate the optical band gap accurately because they were outside the measurable limit. At this point, we observed that the crystal shows a reflection of the light shed from the top, which could be an indication of a possible transition from the insulator to semi-metallic state (Supplementary Fig. 2). Figure 1c also shows the pressure evolution of the band gap obtained using DFT calculations. Although, the optical band gap does not exactly match the electronic band gap and that DFT calculations often underestimate the band gap energies, they can provide a good approximation. Indeed, the calculated values agrees well with the extrapolation of the extracted optical band gap energies from this experiment.

Above 40.0 GPa, a sudden change in the pressure-dependence of the absorption edge was observed, which moves back to a shorter wavelength (Supplementary Fig. 3). The blue shifts of the absorption edge are ascribed to the reopening of the band gap, which are accompanied by a change in the crystal optical color from black to brown and disappearance of the reflected light. The reopening of band gap continues up to 60.0 GPa, where the estimated optical band gap is 2.53 eV, larger than that at ambient conditions (Supplementary Fig. 4). The crystal optical color and the band gap energy remain the same during the pressure release down to 13.0 GPa, when the pressure is completely released, the color of the crystal turned into opaque black with the estimated band gap energy of 1.72 eV. The crystal optical color changes to brown and the band gap reopening are clear indications of loss of conductivity and possible transformation to a new insulating state.

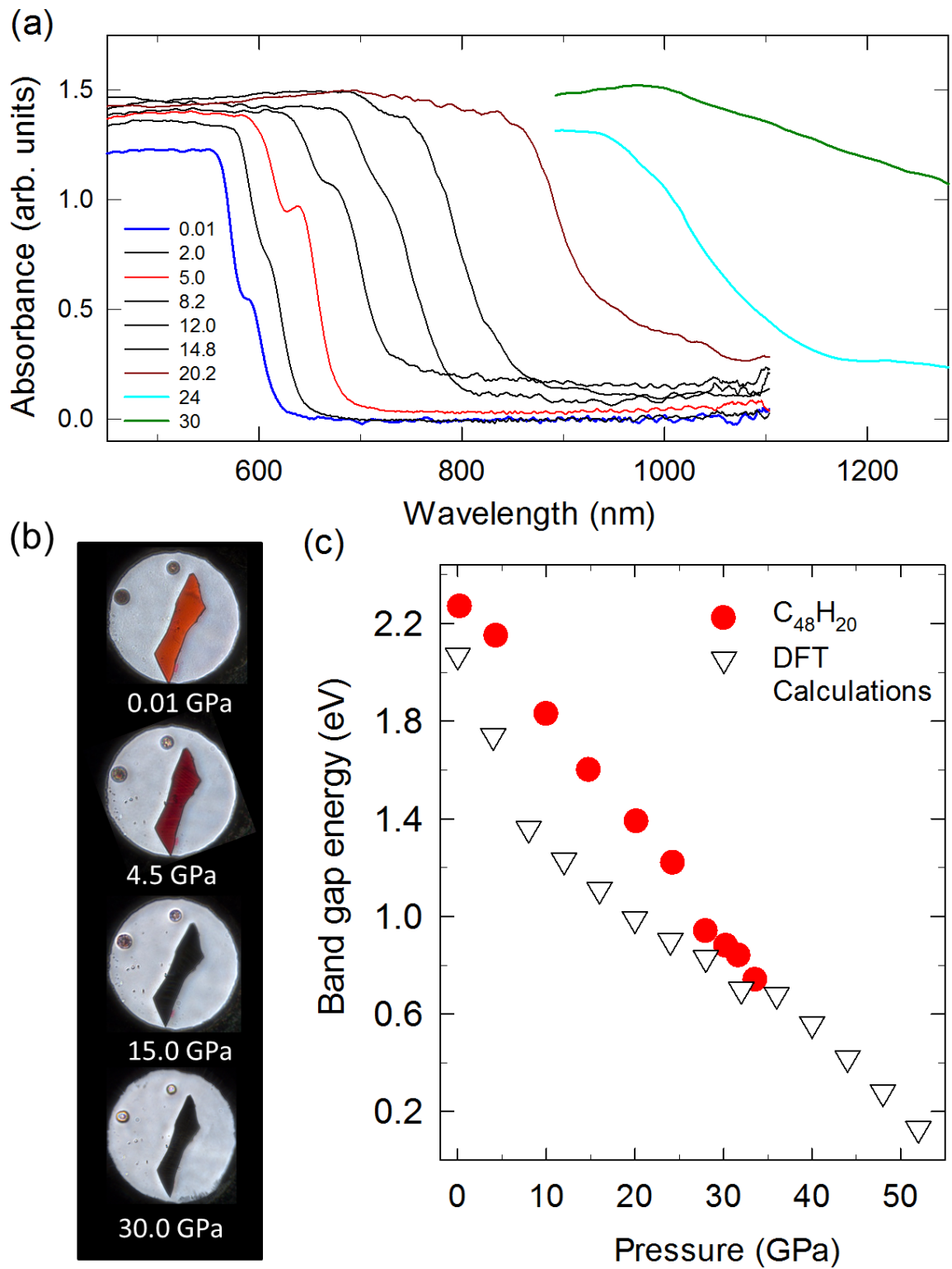

**Fig. 1 In-situ pressure-dependent absorption spectra of a single crystal $C_{48}H_{20}$. a** Optical absorption spectra recorded during compression from ambient pressure up to 33.6 GPa. **b** Optical images of $C_{48}H_{20}$ at each pressure points where the crystal color changes from red, dark-red, opaque black and black. **c** Estimated and calculated band gap energy with respect to pressure.

# The insulator to semiconductor transition

The effects of pressure on the electronic properties and possible pressure-induced electronic transitions of $C_{48}H_{20}$ was further investigated by electrical transport measurements. At ambient conditions, $C_{48}H_{20}$ is an electrical insulator with the resistivity value beyond the sensitivity of the equipment used. With increasing pressure in a diamond anvil cell (DAC) at RT, we observe a pronounced decrease in its electrical resistance (Fig. 2). The resistance was calculated by the Van de Pauw method using the equation:

$$\exp(-\pi R_1/R_s) + \exp(-\pi R_2/R_s) = 1 \tag{1}$$

where $R_1$ and $R_2$ are the two resistances measured by the four-probe method and $R_s$ is the sheet resistance.[34] Electrical resistivity $\rho$ was calculated as $\rho = R_s \times t$, where $t$ is the thickness of the crystal (approximately 10 μm for the typical $C_{48}H_{20}$ crystal). Up to a pressure of about 15.0 GPa, the resistivity gradually decreases from ambient pressure with increasing pressure. Above 15.0 GPa, the rate of resistance decrease increases rapidly, with the resistivity value dropping by three orders of magnitude up to a pressure of about 30.0 GPa. During compression, the crystal optical color was observed to change from red to black at 25.0 GPa. Such a drop in resistance and crystal color change indicate that $C_{48}H_{20}$ undergoes a pressure-induced insulator-semiconductor transition above ~25 GPa. In order to confirm the transition, the temperature dependent resistance measurements were done at three pressure points, 13.0, 23.0 and 38.0 GPa, as shown in Figure 2b. The electrical resistance at 13.0 GPa increases immediately with decreasing temperature, which is a typical insulator character. The electrical resistance at 23.0 GPa, on the other hand, exponentially increases in resistance with decreasing temperature, which is a typical semiconducting character, confirming the insulator-to-semiconductor transition. A rapid increase in RT resistance value was observed above 40.0 GPa (Supplementary Fig. 5), corresponding to a pressure-induced amorphization and agree with band gap opening observed in UV-vis absorption measurement at high pressure.

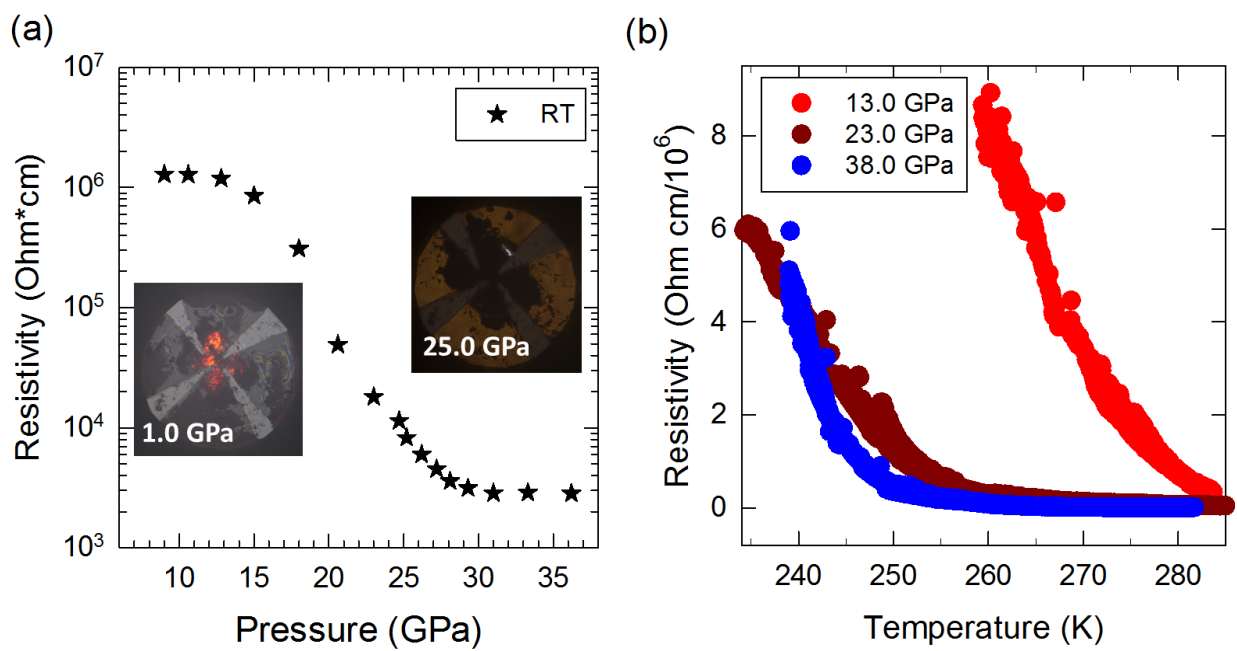

**Fig. 2 Electrical transport properties of $C_{48}H_{20}$ under pressure.** **a** RT resistivity of $C_{48}H_{20}$ at various pressure points. **b** Temperature-dependent resistivity of $C_{48}H_{20}$ obtained during warming from 200 K to 290 K with a van der Pauw setup at 13.0, 23.0 and 38.0 GPa.

### Evolution of crystal lattice vibrational modes at high pressure

The band gap energy and resistivity of the organic materials are closely related to the intermolecular interactions, and Raman spectroscopy is one of the most powerful and informative techniques for monitoring physical changes. To confirm the effects of compression on intermolecular interactions, we carried out an in-situ high-pressure Raman spectroscopy experiment at RT up to 49.0 GPa. Raman spectra were obtained with a laser excitation at 532 nm with a normal power density of 0.5 mW/μm² to avoid spectral changes due to radiation damage. At ambient pressure and below 3.0 GPa, all Raman modes are obscured by the strong fluorescence from the sample, consistent with previous work.[27] Figure 3 shows the evolution of the Raman modes in $C_{48}H_{20}$. The intensity of the entire Raman modes decreases at 23.0 GPa, which coincides with the change in crystal color to complete black. At 26.0 GPa, significant changes in the lattice modes were observed, where the two lattice modes split into two doublets together with the appearance of new modes, implying enhancement of intermolecular interactions with possible structural transitions to lower symmetry. Finally, at 40.0 GPa, all Raman modes disappear, indicating the occurrence of irreversible chemical reaction or amorphization of the molecule. The Raman modes reappear when decompressed from 49.0 GPa, however, the original phase was not recovered even when the pressure was completely released.

To investigate further the effects of pressure on $C_{48}H_{20}$ and their reversibility, we carried out IR measurements up to a maximum pressure of 39.0 GPa and decompressed to ambient conditions. (Supplementary Fig. 6) After decompression from 39.0 GPa, an almost complete recovery of the original spectral features was observed at 0.2 GPa, which also supports that the $C_{48}H_{20}$ was not decomposed into the amorphous state during phase transition above 23.0 GPa but the irreversible chemical reaction sets in above 40.0 GPa.

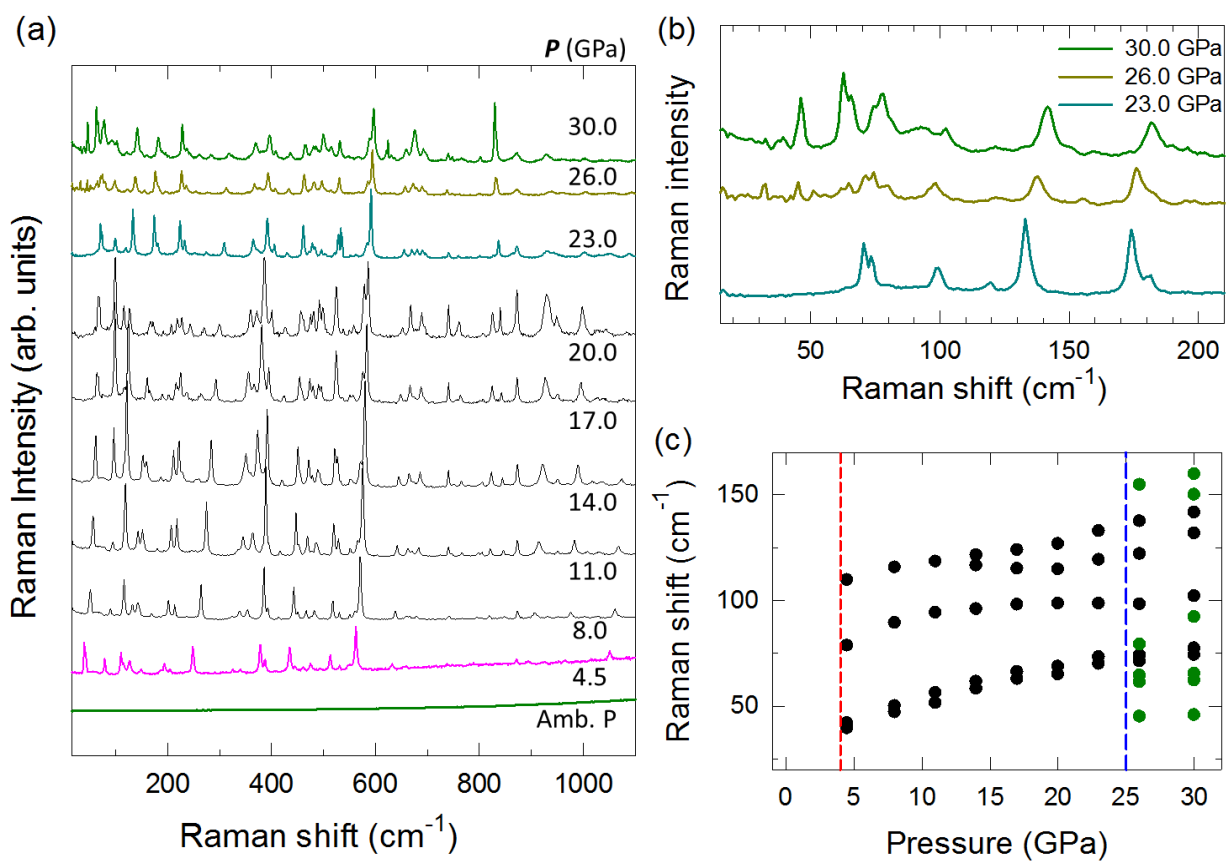

**Fig. 3 High-pressure Raman spectroscopy measurements**. **a** Pressure-dependent Raman spectra of a single-crystal of $C_{48}H_{20}$ recorded during compression with a 532 nm excitation laser. **b** Selected Raman spectra in the low-frequency region of $C_{48}H_{20}$ at pressures above 20.0 GPa. New Raman modes appeared above 23.0 GPa, indicating structural transition to new polymorph. **c** Pressure-dependence of Raman shifts. Raman modes were not observed between ambient pressure and 4.0 GPa due to fluorescence (indicated by red dotted line). Phase transition was observed above 23.0 GPa (blue dotted line).

## Structural evolution under high pressure at RT

To gain insight into structure-to-property relationships under high pressure and to address the structural stability of $C_{48}H_{20}$, we carried out an in-situ high pressure synchrotron powder X-ray diffraction (SPXRD) and SCXRD experiments reaching pressures of up to 60.0 GPa and 26.2 GPa, respectively. Figure 4a depicts selected SPXRD profiles of polycrystalline $C_{48}H_{20}$ obtained at various pressure points. At 0.3 GPa, the Le Bail fit of the SPXRD data shows that the reflection intensities roughly correlate to the single-crystal data obtained at ambient pressure (Fig. 4b, top). Under ambient conditions, $C_{48}H_{20}$ has a monoclinic $\beta$-herringbone structure (Fig. 4c, left panel) with space group $P2_1/c$, which encompasses two translationally inequivalent molecules per unit cell centered on positions 0,0,0 and ½, ½, 0. The $C_{48}H_{20}$ molecules, in particular, are arranged in layers parallel to the

*bc* plane, with the long molecular axis roughly perpendicular to the plane of the layer, whilst neighboring molecules within a layer are twisted with respect to each other, forming a distinctive herringbone stacking. Figure 4c, left panel depicts the projection of $C_{48}H_{20}$ onto the *bc* plane, showcasing two such layers and the arrangement of the molecules inside those layers. Although there are alternations in intensity ratios the (100) and (102) Bragg reflections, structure transition occurs at pressures beyond 18.7 GPa. This transformation is observed by the splitting of the diffraction peak at 4° $2\theta$, which suggests that a reduction in symmetry has taken place. Our best attempts to index the resulting diffraction data yields a unit cell having the centrosymmetric triclinic symmetry with space group *P*-1 (Fig. 4a and Fig. 4b, bottom). The lowering of the symmetry from monoclinic to triclinic would be in accordance with symmetry breaking and group-sub group relations (Fig. 4d). The peak position extraction from an indexed powder diffractogram was used to evaluate the unit cell metrics of polycrystalline data at all pressure points. Supplementary file contains a summary of $C_{48}H_{20}$ unit cell metrics acquired from the Le Bail refinement against SPXRD data ($\lambda$= 0.6199 Å) at RT (see supplementary Table S2). Despite the change in symmetry, the arrangement of the molecules within the unit cell for these two phases is likely to be fairly similar. The correlations between the cell parameters utilized in the literature[35,36] and the characteristic distances (*h*, *x*, and *y*) are shown in Figure 4c, right panel. The phase transition at 23.0 GPa agrees well with the insulator-to-semiconductor transition observed in UV-vis absorption, electrical transport, and Raman measurements.

At pressures higher than 38.0 GPa, the combination of the size and strain-induced effects led to an obvious further broadening of the observed X-ray diffraction peaks, gradual fading away of the triclinic (001) and (011) Bragg peaks and the emerging of amorphous-like broad peaks. When the pressure exceeded 62.0 GPa, the diffraction peaks from the sample become barely distinguishable from the background noise. But, when the pressure is released to ambient conditions, there is a partial recovery of the initial monoclinic phase, accompanied by the reappearance of the (002), (100), (102), and (004) reflections. Yet, reduced intensities and broader peak widths of decompressed monoclinic Bragg reflections point to a substantial level of topological disorder in the relaxed sample.

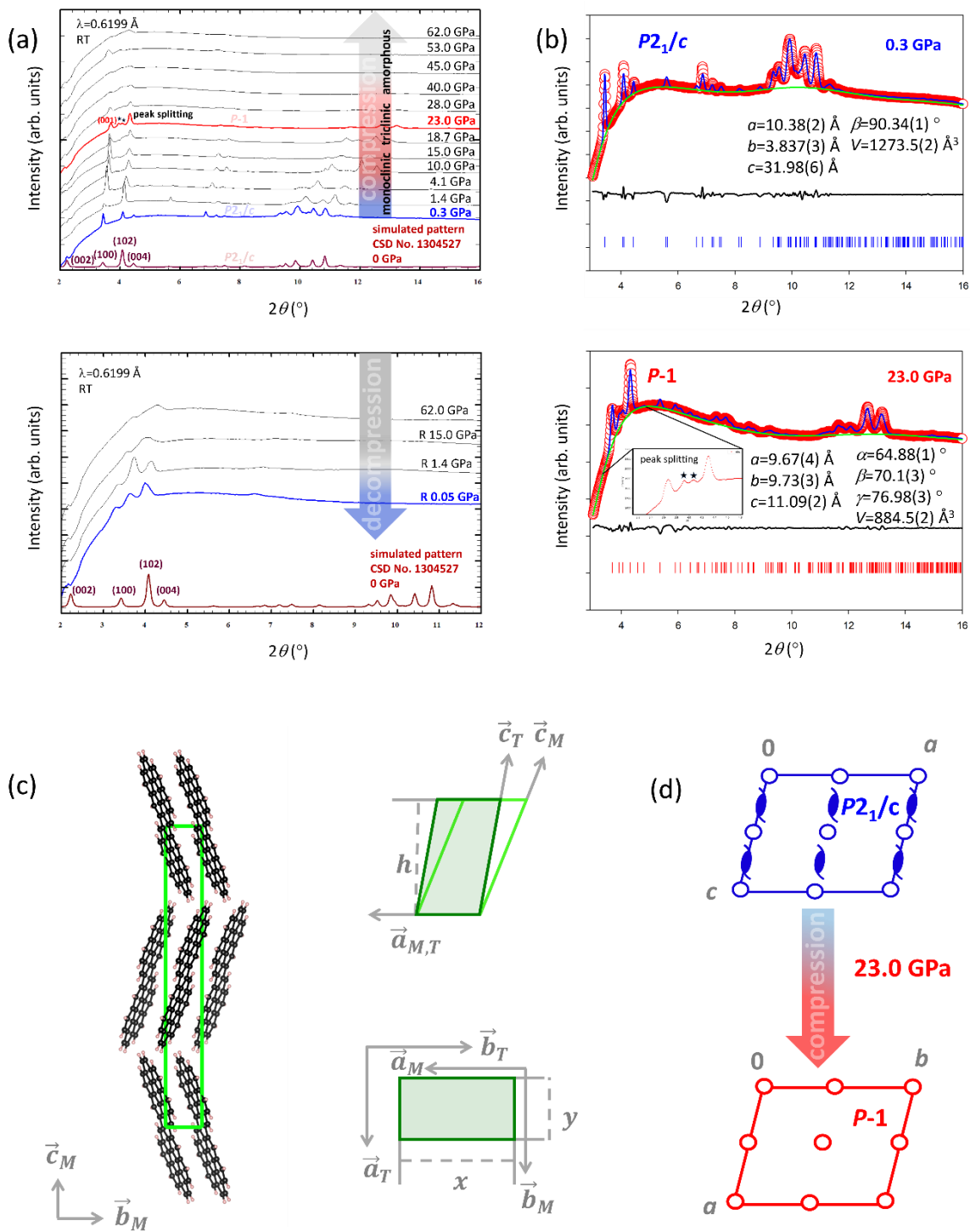

**Fig. 4 High-Pressure SPXRD measurements and structure analysis of $C_{48}H_{20}$. a** Selected in-situ high-pressure SPXRD profiles upon compression (top) and decompression (bottom). The experimental data at 0.3 GPa match the simulated SPXRD pattern of $C_{48}H_{20}$ collected at ambient conditions (CSD No. 1304527).[28] **b (a)** Observed (red circles) and calculated (solid blue line) SPXRD profiles obtained from the Le Bail fit against the PXRD data of $C_{48}H_{20}$ at 0.3 GPa (top) and 23.0 GPa (bottom). The lower solid black line depicts the difference profiles. The solid blue and red vertical tick marks show the reflection positions of the monoclinic and triclinic $C_{48}H_{20}$ phases, respectively. The lower solid green lines indicate the fitted background contribution. **c** Crystal packing of $C_{48}H_{20}$ at ambient conditions, viewed along *a* axis (left panel). Cell conventions for oligoacenes as used in literature[36,37] (right panel). The index M stands for monoclinic cell, and T for triclinic cell. The characteristic distances are the layer thickness, *h*, and the long and short herringbone lengths *x* and *y*. **d** Schematic projection of the equatorial plane of point groups 2/*m* in the monoclinic

phase and -1 in the triclinic phase during the structural transition triggered by hydrostatic pressure, in accordance with symmetry breaking and group-sub group relations.

To obtain more comprehensive data on the molecular structure and evolution of periodic arrangement in three-dimensional space of $C_{48}H_{20}$, we collected SCXRD data at pressures of 3.3, 10.3, 19.2, 26.2 GPa. The refined unit cell parameters from these measurements are listed in Table S3, where the structures were solved as monoclinic crystals in the $P2_1/c$ space group and the crystallographic information on the full structure solution and refinement in Tables S4-S7. An anisotropic response to compression along three crystallographic axes was also found, with the lattice parameter $b$ reduced by 12.8% at 26.2 GPa, while $a$ and $c$ were reduced by 4.2% and 4.0%, respectively, yielding a volume reduction of 20.0%. The intermolecular distances were also calculated based on the experimental data, where at ambient pressure the C-C and H-H distances between atoms on nearest-neighbor molecules were 3.474(3) Å and 2.354(1) Å, respectively. At 26.2 GPa, SCXRD data analysis shows the same C-C and H-H intermolecular distances were reduced to 2.724(7) Å and 1.87(7) Å. The structure models, the pressure-volume datapoints, and lattice parameters obtained SCXRD and from DFT calculations showed good agreement to one another. The crystal appearance changed significantly when compressed from 19.2 to 26.2 GPa. Data collection immediately after pressure increase to 26.2 GPa enabled a good structure model, consistent with those obtained at lower pressures. However, data collected on the same crystal at the same pressure approximately 40 minutes later revealed a pronounced degradation of the crystal quality (Supplementary Fig. 7), with the majority of reflections corresponding to the crystal having almost disappeared, and those remaining being streaky and of very weak intensity. This suggests a progressive structural phase transition to take place at this pressure, consistent with the transformation observed from SPXRD between 18.7 and 23.0 GPa.

Figure 5 depicts evolution of the unit cell volume, electron localization function (ELF), and band structures of $C_{48}H_{20}$ molecules. The pressure-volume datapoints obtained from SPXRD, SCXRD and DFT calculations are almost identical to one another while the structure remained monoclinic in the $P2_1/c$ space group (Fig. 5a). A third-order Birch-Murnaghan equation of state was used in the EoSFIT software[37] to fit the pressure-volume datapoints for SPXRD ($V_0$ = 1241.4(2) Å$^3$, $K_0$ = 9.0(3) GPa) and for SCXRD ($V_0$ = 1259.6(2) Å$^3$, $K_0$ = 9.1(2) GPa). The ELF calculations based on the structural information at ambient pressure and 26.2 GPa (Fig. 5b) show that the van der Waals space between $C_{48}H_{20}$ is reduced as the intermolecular distances were compressed and with rotation of molecules, however, expected intermolecular charge density overlap was not observed. Electronic density of states (eDOS) calculations, also based on the solved crystal structures at various pressure points, were performed and suggest that band gap of $C_{48}H_{20}$ narrows rapidly with pressure, and even becomes a poor metal at 28.0 GPa (Fig. 5c).

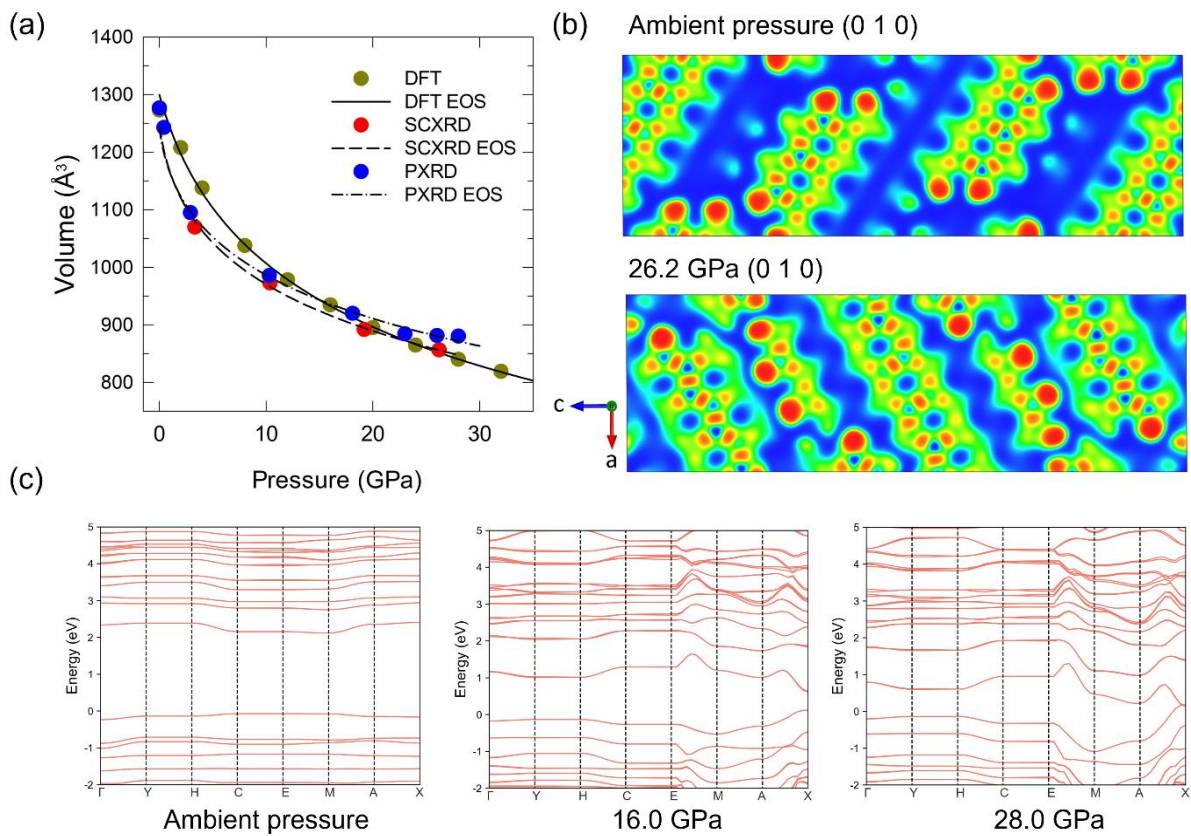

**Fig. 5 Equation of states and electronic structures. a** The P-V equation of states based on high-pressure experiments and DFT calculation. The solid and dashed lines represent the third-order Birch-Murnaghan equation of state fitted to the volume data. **b** The charge density distribution of $C_{48}H_{20}$ at ambient pressure and 26.2 GPa calculated from crystal structures obtained from SCXRD. **c** Selected band structures at 1bar, 16.0 and 28.0 GPa. The HOMO and LUMO come into contact at 28.0 GPa. The distance between high-symmetry points is normalized for clarity purposes.

During compression, $C_{48}H_{20}$ molecules are rearranged in terms of distances and angles in their herringbone motif, resulting in a gradual darkening of optical crystal colors and an enhancement in intermolecular interactions. We observed $C_{48}H_{20}$ to undergo a phase transition around 23.0 GPa with a slightly different crystal structure, where the exact pressure probably depends on the hydrostaticity on the sample. The underestimation of the band gap and metallization pressure could be due to not taking this rearrangement of the molecules into account.

## Discussion

PAHs are the simplest and most abundant organic molecules, which is also known as a strong electron-phonon coupled systems. The magnitude of their electron-phonon interaction depends on the distance between the nearest neighbour molecules[38] and thus, the knowledge of the intermolecular distance can be crucial in understanding the mechanism for physical properties under high pressure. However, detailed information on the crystal structure, orientation of the molecules, atomic positions, and the relative distance at pressures above 10 GPa of any PAHs are extremely limited.[39] From our optical and structural data, the overall scenario is as the nearest-neighbor C-C distance for $C_{48}H_{20}$ approaches 2.7 Å under pressure, the intermolecular interactions are enhanced and the optical band gap decreases rapidly up to 34.0 GPa. The molecular units remain stable at least up to 39 GPa, and from the equation of state fitting, the C-C distance is expected, assuming that the compressibility remains the same between different molecular arrangements, to become 2.6 Å at around 40.0 GPa (Supplementary Fig. 8). Above 40.0 GPa, we observed the reopening of the band gap energy, increase in the RT resistivity, and disappearance of Raman modes. It is fair to predict that the C-C distance between 2.6 and 2.75 Å is critical distance for charge density overlap as expected in the pressure range of 23.0-38.0 GPa. On further compression, chemical reactions will take place. This is in agreement with the critical intermolecular distances suggested by previous molecular-dynamics simulations of $C_6H_6$ at 23.0 GPa and 540 K that a potential energy barrier occurring at 2.7 Å intermolecular separation must be overcome for intermolecular chemical bonding to occur, while the reaction threshold for irreversible reactions initiate when the nearest neighbor C-C distance approaches 2.5-2.6 Å.[22] Further IR spectroscopy measurements have indicated that the formation of amorphous hydrogenated carbon or carbon nanothreads commenced at pressures higher than 16.2 GPa for $C_6H_6$.[40,41]

We have also demonstrated that the band gaps of PAH molecules can be chemically pre-compressed without the addition of other elements. The optical band gap of benzene at ambient pressure is 6 eV, which can be chemically compressed down to 2.2 eV (2.01 eV by DFT simulation) by fusing benzene rings to form a fully benzenoid structure. In addition, they exhibited significantly high chemical stability, with an irreversible chemical reaction only occurring at pressures around 40 GPa. Physical compression reduced the band gap of $C_{48}H_{20}$ from 2.2 eV at ambient conditions to 0.7 eV at 34.0 GPa and, above 23.0 GPa, there was an insulator to semiconductor transition. It should be noted that this single-component single-phase single-crystal of pre-compressed large PAH molecules is stable at ambient conditions and can be fully characterized prior to the application of physical compression. This point is a huge advantage over common so-called superhydrides materials, which require high-pressure high-temperature synthesis in DAC resulting in multi-component multi-phased crystals in micron scale.[42,43]

With the aim of exploring superconductivity in a compound consisting of only carbon and hydrogen, we studied the pressure evolution of the electronic and optical properties as well as the crystal structure of $C_{48}H_{20}$. The UV-vis absorption measurements and the electrical resistance measurements show a pressure-induced transition from insulator to semiconductor, with the semiconducting behavior observed in the pressure range between 23.0 to 38.0 GPa. The abrupt phase transition was coupled with a change in crystal color from red to black, a narrowing of the band gap below 1.0 eV at 24 GPa, and an increase in intermolecular interaction. We discovered the first hydrocarbon that undergoes an insulator-to-semiconductor transition at a pressure well below megabar. Our discovery promotes further exploration of the pressure responses of the other large PAH molecules to a possible transition to the metallic and superconducting states.

# Experimental Section

***Sample Preparation and Characterization***. The single-crystal dicoronylene ($C_{48}H_{20}$) was obtained by the vapor phase fusion reaction of coronene ($C_{24}H_{12}$) purchased from TCI chemical (>95%). The details of crystal preparation, characterizations under ambient conditions, and structure-to-optical properties relationships below 20 GPa are described in recent work.[27]

***High-Pressure Electrical Measurements***. The electronic transport properties under high pressure and low temperature were investigated via van der Pauw electrical conductivity method in a symmetric diamond anvil cell (DAC). The pressure was generated by a pair of diamonds with a 300 μm diameter culet. A rhenium gasket was pressed and drilled hole at the center of the gasket with a diameter of the diamond culet. A cubic boron nitride (c-BN)-epoxy insulation layer was prepared to protect the electrode leads from the metallic gasket. 150 μm diameter hole was drilled as a sample chamber at the center of the c-BN to place the $C_{48}H_{20}$ sample with NaCl as a pressure transmitting medium (PTM). Four platinum strips were arranged to contact the sample in the chamber. The first-round experiments at different pressure-temperature conditions were performed using a laboratory-designed (with optical window) electrical transport system where a Keithley 6221 current source, 2182A nanovoltmeter, and 7001 switch device were used as the current supply, voltmeter, and voltage/current switcher, respectively. The pressure at the sample chamber was measured using the ruby and diamond fluorescence method. The second-round high-pressure electrical transport experiment was conducted using the Quantum Design physical property measurement system using 23 mm-type CuBe cells.

***Optical Spectroscopy***. Measurements of UV-vis absorption and optical images were performed with in-house developed Gora-UVN-FL (Ideaoptics, Shanghai) micro-region spectroscopy system. A single-crystal dicoronylene was loaded together with ruby ball into the sample chamber of a Mao-type symmetric DAC with 400 μm culet-sized low-fluorescent anvils. Si-oil was used as a PTM. Each pressure point was determined by the fitting pressure shift of the ruby fluorescence line.

***First-Principles Calculations.*** First-principles calculations are implemented under the framework of density functional theory using the Vienna's ab initio simulation package (version 6.3.4). We employ the generalized gradient approximation (GGA) with the parameterization of Perdew-Burke-Ernzerhof optimized for solids.[44,45] The valence electrons of C ($2s^2 2p^2$) and H ($1s^1$) are decided by the projected augmented wave potentials. The cut-off kinetic energy for plane-wave basis set is 700 eV, which is found sufficient to converge the force acting on each atom less than 0.01 eV/Å. The unit cells of $C_{48}H_{20}$ (96 C and 40 H) are used for calculating electronic structures and their Brillouin zone is sampled by a Monkhorst mesh with 0.2 $A^{-1}$ as the spacing between k points. We approximate the long-range pair interactions using a simple D2 method of Grimme,[46] with cut-off radius of 50 Å. At each target pressure, the unit cell is optimized for atomic position, cell shape and cell volume. After structural optimization, we performed one self-consistency cycle with the Heyd-Scuseria-Ernzerhof screened hybrid function (screening factor 0.15)[47] to obtain the electronic structures, as shown in Figure 5.

*Vibrational Spectroscopy*. High-pressure Raman measurements were performed with a custom-built highly focused Raman system, using 532 nm laser source. A single-crystal dicoronylene was loaded along with a ruby ball into the sample chamber of a Mao-type symmetric DAC with 400 μm culet-sized low fluorescence anvils. Si oil was used as a PTM. Measurements were imaged on a CCD, with a typical sample exposure of 0.01s at 0.5 mW incident laser power. High-pressure Mid-IR spectra were obtained with Bruker VERTEX 70v IR spectrometer with HYPERION 2000 microscope in transmission mode in the range of 600-5000 $cm^{-1}$ in 2 $cm^{-1}$ per step. The spectrum of an empty DAC at ambient pressure was used as the background signal for all measurements. The sample chamber was filled with dicoronylene single crystals and compressed without PTM.

*Synchrotron Powder X-Ray Diffraction (SPXRD) Measurements*. High pressure synchrotron PXRD measurements were performed at BL12B2 at SPring-8 (Japan; λ = 0.6199 Å, beam diameter of 50 μm). Polycrystalline $C_{48}H_{20}$ was loaded along with a ruby ball into the sample chamber of a Mao-type symmetric DAC with 400 μm and 200 μm culet-sized anvils. The sample chamber was fully packed with the sample and Daphne oil was used as a PTM and each pressure point was determined by the fitting pressure shift of the ruby fluorescence line. Twenty 60-second scans were acquired and summed to ensure a good signal-to-noise ratio while avoiding the possibility of radiation damage to the sample. The two-dimensional images were integrated with IPAnalyzer and data analysis was performed using the LeBail pattern decomposition technique within the GSAS-II program suite.[48]

*Synchrotron Single-Crystal X-Ray Diffraction (SCXRD) Measurements*. A single crystal of $C_{48}H_{20}$ was screened in-house to check diffraction quality prior to loading. A well-diffracting crystal of approximate dimensions 42 x 15 x 10 $μm^3$ (LxWxH) was loaded with 2000 bars of He gas as the pressure transmitting medium in a BX90 DAC using diamond anvils with 250 μm culets and a Re gasket. A small (diameter ≈ 3 μm) grain of gold, and a ruby sphere (diameter ≈ 5 μm) were included for pressure determination. Collections of single-crystal X-ray diffraction (SCXRD) data were performed at the pressure post gas loading (3.3 GPa), as well as at 10.4, 19.2, and 26.2 GPa, stopping once the sample was no longer diffracting sufficiently for crystal structure determination.

These SCXRD experiments were performed at the ID15b beamline of the European Synchrotron Radiation Facility (ESRF) in Grenoble, France. Diffraction experiments used a monochromatic X-ray beam of wavelength 0.4099 Å and beam spot size of approximately 1.0 x 1.0 μm$^2$, data were collected on a Dectris Eiger2 X CdTe 9M detector in 0.5° angular steps over a -36 to 36° range. Data were processed using the CrysAlis$^{Pro}$ software[49] for reflection search, removal of parasitic diamond anvil reflections, unit cell determination, and data integration. Structure solution and model refinement were performed using the Olex2 user interface.[50] Initial structure solution of the first data set (the lowest pressure point) was achieved with intrinsic phasing methods using SHELXT.[51] For the subsequent datasets, the model from the prior pressure point was used as a starting point for least squares refinements of the structure using SHELXL.[52] Hydrogen atoms were constrained to calculated geometries and allowed to ride their parent atoms.

**Data availability**

All relevant data and single crystal samples are available from the corresponding author.

The data that support the findings of this study are available in the Supporting Information. The CCDC number of the dicoronylene, $C_{48}H_{20}$, at ambient pressure is 2141267, at 3.3 GPa is 2352043, at 10.4 GPa is 2352046, at 19.2 GPa is 2352045, and at 26.2 GPa is 2352044.

**Acknowledgments:** This research was financially supported by the Beijing Natural Science Foundation (Funding No. IS24025), the National Key Research and Development program of China (Grant No. 2022YFA1402301 and 2022YFA1405500) and the National Natural Science Foundation of China (Grant No. U2230401). D.L. thanks the UKRI Future Leaders Fellowship (MR/V025724/1) for financial support. The SPXRD experiments at BL12B2 at the SPring-8 were performed under SPring-8 proposal No. 2023A4139. The authors acknowledge the European Synchrotron Radiation Facility (ESRF) for the provision of beamtime at the ID15b beamline. For the purpose of open access, the author has applied a Creative Commons Attribution (CC BY) licence to any Author Accepted Manuscript version arising from this submission.

**Conflict of interest**

The authors declare no conflict of interest.

**Contributions**

T.N. contributed to conceptualization, funding acquisition, preparing sample, investigations, formal analysis, writing the paper. T.N. K.B. and X.L. performed absorption spectroscopy measurements and analysis. T.N., C.Z. and K.B. performed transport property measurements, T.N. and P.D-S. Performed Raman and IR measurements, T.N., H.I., and N.H. performed high pressure SPXRD. M.V. performed formal analysis of the diffraction data and review&editing the paper. S.B., D.L., A.L., A.R. performed high pressure synchrotron SCXRD and formal analysis of the diffraction data. Q.H. performed all DFT calculations and review&editing the paper. H-k.M. and Y.D. supervised the study, funding acquisition, and review&editing the paper. All authors discussed the results and made substantial input to the writing.